\documentclass[aps,prc,twocolumn,superscriptaddress]{revtex4-1}

\usepackage{graphicx}
\usepackage{amsmath}
\usepackage{braket}
\usepackage{url}
\usepackage{multirow}
\usepackage{multirow}
\usepackage{amssymb}
\usepackage{booktabs}
\usepackage{rotating}
\usepackage{bm}
\usepackage{bbm}
\usepackage{ulem}
\usepackage{textcomp}

\usepackage{verbatim}

\usepackage[colorlinks,
linkcolor=blue,
anchorcolor=blue,
urlcolor=blue,
citecolor=blue]{hyperref}

\usepackage[mathlines]{lineno}

\begin{document}


\newcommand{\rwYQ}[1]{ {\color{green} #1}}

\title{Complex valence-space effective operators for observables: The Gamow-Teller transition}

\author{Z. C. Xu}
\affiliation{
School of Physics, and State Key Laboratory of Nuclear Physics and Technology, Peking University, Beijing 100871, China}
\author{S. Zhang}
\affiliation{
School of Physics, and State Key Laboratory of Nuclear Physics and Technology, Peking University, Beijing 100871, China}
\author{J. G. Li}
 \affiliation{%
Institute of Modern Physics, Chinese Academy of Sciences, Lanzhou 730000, China
}%
\affiliation{Southern Center for Nuclear-Science Theory (SCNT), Institute of Modern Physics, Chinese Academy of Sciences, Huizhou 516000, Guangdong Province, China}
 \affiliation{
School of Physics, and State Key Laboratory of Nuclear Physics and Technology, Peking University, Beijing 100871, China}

\author{S. L. Jin}
 \affiliation{
School of Physics, and State Key Laboratory of Nuclear Physics and Technology, Peking University, Beijing 100871, China}
\author{Q. Yuan}
 \affiliation{
School of Physics, and State Key Laboratory of Nuclear Physics and Technology, Peking University, Beijing 100871, China}

\author{Z. H. Cheng}
\affiliation{
School of Physics, and State Key Laboratory of Nuclear Physics and Technology, Peking University, Beijing 100871, China}

\author{N. Michel}%
 \affiliation{%
Institute of Modern Physics, Chinese Academy of Sciences, Lanzhou 730000, China
}%
\affiliation{Southern Center for Nuclear-Science Theory (SCNT), Institute of Modern Physics, Chinese Academy of Sciences, Huizhou 516000, Guangdong Province, China}

\author{F. R. Xu}%
\email{frxu@pku.edu.cn}
\affiliation{
School of Physics, and State Key Laboratory of Nuclear Physics and Technology, Peking University, Beijing 100871, China}
\affiliation{Southern Center for Nuclear-Science Theory (SCNT), Institute of Modern Physics, Chinese Academy of Sciences, Huizhou 516000, Guangdong Province, China}

\begin{abstract}
Nuclei in the vicinity of driplines have been receiving a lot of attention in nuclear structure studies. In the nuclei, the continuum coupling is crucial in reproducing weakly bound and unbound phenomena. To calculate observables of the nuclei as open quantum systems, we have developed valence-space effective operators in the complex-energy Berggren basis using many-body perturbation theory. We focus on the Gamow-Teller $\beta$ decay in the {\it sd} shell. The two- plus three-nucleon force from the chiral effective field theory, named EM1.8/2.0, has been used. The Gamow shell model which takes the continuum coupling into account can properly reproduce experimental observations of weakly bound and unbound states. The $\beta$-decay isospin asymmetry between the dripline nucleus $^{22}\rm Si$ and its mirror partner $^{22}\rm O$ is reproduced, in which the $s_{1/2}$ continuum plays a key role. Significant Thomas-Ehrman shift is seen through mirror energy differences between the mirror daughters $^{22}\rm Al$ and $^{22}\rm F$, in which the continuum effect plays an important role.
\end{abstract}{}

\maketitle
One of the frontier issues in nuclear physics theory is the description of weakly bound and unbound nuclei in the vicinity of driplines. The development of rare-isotope beam facilities worldwide is being driven primarily by this topic. These exotic nuclei are important for many longstanding problems, such as the limits of the nuclear landscape~\cite{SAKURAI1999180,PhysRevLett.105.032501,nature.486.509,PhysRevLett.126.022501,PhysRevLett.123.212501}, the formation and evolution of new shell closures~\cite{nature.498.346,nature.502.207,nature.569.53}, and astrophysical nucleosynthesis~\cite{MUMPOWER201686,PhysRevLett.116.121101}. Isospin asymmetry happens with significant mirror energy difference (MED) typically. The Thomas-Ehrman shift (TES)~\cite{PhysRev.88.1109,PhysRev.81.412} can occur when the nuclear state in the proton-rich mirror nucleus is weakly bound or unbound, especially for the states with a significant $s$ partial wave and hence a strong coupling to the continuum~\cite{ZHANG2022136958}.

The goal of the {\it ab initio} nuclear theory is to describe the structures of nuclei from the underlying interactions between nucleons without input from experimental data beyond that necessary to implement nuclear forces. Using two- and three-nucleon forces (2NF and 3NF, respectively) from the chiral effective field theory (EFT)~\cite{RevModPhys.77.427,MACHLEIDT20111,PhysRevLett.99.042501}, {\it ab initio} many-body calculations have exhibited great progresses, such as in reproducing the location of the oxygen dripline~\cite{PhysRevLett.105.032501,PhysRevLett.110.022502}, and understanding the origin of the anomalous long lifetime of $^{14}\rm C$~\cite{PhysRevLett.106.202502}. A variety of observables of nearly all open-shell nuclei that are accessible to the conventional shell model (SM) can be obtained using the many-body perturbation theory (MBPT) method~\cite{KUO197165,10.3389/fphy.2020.00345}, including energies, charge radii, electromagnetic moments and transitions, and $\beta$ decays~\cite{PhysRevC.95.064324,PhysRevC.100.014316}. 

Since nuclei around driplines are open quantum systems, the continuum coupling and resonance degrees of freedom are crucial in describing their structure. The transition density between initial and final states is indeed influenced by the asymptotic behaviors of wave functions.
A powerful tool for properly describing the asymptotic behavior of wave functions is the Gamow shell model (GSM)~\cite{Michel2021,PhysRevLett.89.042501,PhysRevLett.89.042502} which uses the complex-energy Berggren basis~\cite{BERGGREN1968265} so that it incorporates the coupling to the continuum at the basis level. 
The complex coupled cluster~\cite{HAGEN2007169,PhysRevLett.108.242501} and complex in-medium similarity renormalization group~\cite{PhysRevC.99.061302} have also been formulated within the Berggren basis to properly account for the continuum coupling.

The Berggren basis can be generated using the complex-energy Gamow Hartree-Fock (GHF) method~\cite{PhysRevC.73.064307,HU2020135206,ZHANG2022136958}. The basis is then produced self-consistently by the used realistic interaction instead of a parameterized Woods-Saxon potential. The GSM MBPT calculations have shown that wave functions incorporating the continuum coupling can provide a suitable description of nuclear states, which can be used to explain phenomena such as the TES~\cite{ZHANG2022136958} and Borromean structure~\cite{MA2020135673}. Therefore, in order to calculate the observables of weakly bound and unbound nuclei more properly, we have developed the theory of valence-space effective operators using MBPT within the Berggren basis. 



Starting from chiral 2NF and 3NF, the intrinsic Hamiltonian of the $A$-nucleon system reads
\begin{equation}
    H = \sum_{i=1}^{A}\left(1-\frac{1}{A}\right)\frac{p_{i}^{2}}{2m}+\sum_{i<j}^{A}\left(v_{ij}^\text{NN}-\frac{\bm{p}_{i}\cdot \bm{p}_{j}}{mA}\right)+\sum_{i<j<k}^{A}v_{ijk}^\text{3N},
    \label{eq1}
\end{equation}
where $\bm{p}_i$ is nucleon momentum in the laboratory coordinate, and $m$ is nucleon mass, while $v^\text{NN}$ and $v^\text{3N}$ are for 2NF and 3NF, respectively. The chiral 2NF plus 3NF labeled by EM1.8/2.0~\cite{PhysRevC.83.031301} has been used, which can globally reproduce nuclear binding energies~\cite{PhysRevLett.126.022501,PhysRevC.105.014302}.

For the {\it ab initio} goal, we use the GHF approximation with the same chiral interaction to generate the Berggren basis which provides bound, resonance and continuum states on equal footing in the complex-momentum (complex-$k$) plane. For the $A\approx 20$ nuclei of interest, $^{16}\rm O$ is chosen as the reference state for the GHF and as the core for the GSM calculation. A brief formulation of the GHF calculation with exact 3NF included can be found in our previous work~\cite{zhang2023ab}.
In practical calculations, the continuum states on the contour $L^{+}$in the complex-$k$ plane need to be discretized, which can be achieved by the Gauss-Legendre quadrature method~\cite{PhysRevC.73.064307,LIOTTA19961}. We use the contour $L^{+}$ with 35 discretization points, which is sufficient to obtain convergence in numerical calculations~\cite{Z.H.Sun-Phys.Lett.B.769.227(2017),HU2020135206,PhysRevC.103.034305}. 
After that, we transform the chiral interaction and bare operator matrix elements from the harmonic oscillator (HO) basis to the GHF basis for the many-body GSM calculation by computing overlaps between the GHF and HO bases wave functions~\cite{PhysRevC.102.034302}. In this calculation, we take the HO basis at $\hbar\omega=16$ MeV with 13 major shells (i.e., $e=2n+l\leq e_{\rm max}=12$) and 
$e_{3\text{max}}=e_1+e_2+e_3\leq12$ for 3NF.

The GHF calculation gives bound $0d_{5/2}$ and resonant $0d_{3/2}$ orbits for both neutrons ($\nu$) and protons ($\pi$). While $\pi 1s_{1/2}$ orbit is a resonance, $\nu 1s_{1/2}$ orbit is bound. The $s_{1/2}$ and $d_{3/2}$ partial waves are treated in the complex-$k$ GHF basis to include the continuum effect, whereas the $d_{5/2}$ partial wave is represented in the real-energy discrete HF basis. The active space for the present GSM calculations are \{$\nu0d_{5/2}$, $\nu1s_{1/2}$ plus continuum, $\nu0d_{3/2}$ resonance plus continuum, $\pi0d_{5/2}$, $\pi1s_{1/2}$, $\pi0d_{3/2}$\} for neutron-rich nuclei and \{$\nu0d_{5/2}$, $\nu1s_{1/2}$, $\nu0d_{3/2}$, $\pi0d_{5/2}$, $\pi1s_{1/2}$ resonance plus continuum, $\pi0d_{3/2}$ resonance plus continuum\} for proton-rich nuclei, respectively.

In many-body calculations, 3NF is usually normal-ordered with respect to a reference state, giving the normal-ordered zero-, one-, and two-body terms with the residual three-body term neglected~\cite{PhysRevLett.109.052501,PhysRevLett.126.022501, PhysRevC.105.014302,ZHANG2022136958, zhang2023ab}.
We construct the valence-particle effective Hamiltonian and other effective operators in the framework of MBPT~\cite{KUO197165} consistently. We separate the Hamiltonian of the reference state into a zero-order part $H_0$ and a perturbative part $H_1$, 
\begin{equation}
    H=H_{0}+(H-H_{0})=H_0+ H_1.
\end{equation}
$H_0$ can take the one-body part of the normal-ordered Hamiltonian, and $H_1$ is the residual two-body part including the normal-ordered 3NF at the two-body level~\cite{ZHANG2022136958, zhang2023ab}. 

For the GSM calculation, valence-space single-particle energies and effective interaction matrix elements can be obtained using so-called $\hat{S}$-box~\cite{CORAGGIO200543} and $\hat{Q}$-box folded diagrams~\cite{SHURPIN197761,Z.H.Sun-Phys.Lett.B.769.227(2017)}, respectively. The $\hat{S}$ box is by definition the one-body part of the $\hat{Q}$ box. Because the GHF basis states with continuum states are not degenerate, we use the extended EKK method~\cite{TAKAYANAGI201161,PhysRevC.89.024313} to construct the effective Hamiltonian $H_\text{eff}$ by iterating
\begin{equation}
\label{H_eff}
    H^{(\kappa)}_{\text{eff}} = PH_{0}P +\hat{Q}(\epsilon)+\sum_{n=1}^{\infty}\frac{1}{n!}\frac{d^n \hat{Q}(\epsilon)}{d\epsilon^n}\{H_{\text{eff}}^{(\kappa-1)}-\epsilon\}^n,
\end{equation}
where $\kappa$ represents the $\kappa$th iteration, and $\epsilon$ is the starting energy. 
The $\hat{Q}$ box is defined as
\begin{equation}
    \hat{Q}(\epsilon)=PH_1P+PH_1Q\frac{1}{\epsilon -QHQ}QH_1P
\end{equation}
with derivatives as
\begin{equation}
\hat{Q}_{n}(\epsilon) = \frac{1}{n!} \frac{d^{n}\hat{Q}(\epsilon)}{d\epsilon^n},
\end{equation}
where $P$ and $Q$ are projection operators representing the model space and its complementary space (the excluded space), respectively, with $P+Q=1$. 
Usually the $\hat{S}$ box and $\hat{Q}$ box are calculated up to the third order and second order, respectively, within the GHF basis~\cite{Z.H.Sun-Phys.Lett.B.769.227(2017),HU2020135206}. In the present work, we have promoted the complex MBPT calculation with the two-body matrix elements of pole states (i.e., bound and resonant states) calculated up to the third order.

After the $\hat{S}$-box and $\hat{Q}$-box calculations, we obtain the complex GSM effective Hamiltonian~\cite{zhang2023ab} in the chosen valence space with the $^{16}{\rm O}$ core.
The complex-symmetric GSM effective Hamiltonian is diagonalized in the model space using the Jacobi-Davidson method in the $m$ scheme~\cite{MICHEL2020106978}.

For other observables, 
their bare operators also need to be renormalized into the valence space, which can be done by a so-called $\hat{\Theta}$ box within the same complex MBPT framework, similar to the $\hat Q$ box. A valence-space effective operator, denoted by  $\Theta_{\text{eff}}$, which takes into account the contribution from the excluded $Q$ space, can be expressed as
\begin{equation}
\Theta_{\text{eff}}=\sum_{\alpha, \beta}|\psi_{\alpha}\rangle\langle\tilde{\Psi}_{\alpha}|\Theta| \Psi_{\beta}\rangle\langle\tilde{\psi}_{\beta}|,
\end{equation}
where the valence-space wave function $|\psi_{\alpha}\rangle$ obtained from diagonalizing $H_{\text{eff}}$ is the projection of the full-space wave function $|\Psi_{\alpha}\rangle$ onto the valence space, i.e., $\left|\psi_{\alpha}\right\rangle=P\left|\Psi_{\alpha}\right\rangle$.

In the MBPT, the $\hat{\Theta}$ box is defined as~\cite{10.3389/fphy.2020.00345, 10.1143/ptp/93.5.905}
\begin{equation}
\hat{\Theta}(\epsilon)=P \Theta P +P \Theta Q \frac{1}{\epsilon - QHQ}QH_{1}P
\end{equation}
and
\begin{equation}
\hat{\Theta}(\epsilon_{1};\epsilon_{2}) = PH_{1}Q\frac{1}{\epsilon_{1} - QHQ}Q \Theta Q\frac{1}{\epsilon_{2} - QHQ}Q H_{1} P
\end{equation}
with their derivatives
\begin{equation}
\label{Chi_n}
\hat{\Theta}_{n} = \frac{1}{n!} \frac{d^{n}\hat{\Theta}(\epsilon)}{d\epsilon^{n}}
\end{equation}
and 
\begin{equation}
\label{Chi_mn}
\hat{\Theta}_{mn} = \left.\frac{1}{m!n!} \frac{d^{m}}{d\epsilon^{m}_{1}} \frac{d^{n}}{d\epsilon^{n}_{2}}\hat{\Theta}(\epsilon_{1};\epsilon_{2})\right|_{\epsilon_{1} =\epsilon_{2} = \epsilon}.
\end{equation}
With the identity $\hat{Q}\hat{Q}^{-1}=1$, the final perturbative expansion of the effective operator $\Theta_{\text{eff}}$ can be expressed by the $\hat{Q}$ box and $\hat{\Theta}$ box, as
\begin{equation}
\label{Theta_eff}
\begin{aligned}
    \Theta_{\text{eff}}=&(P+\hat{Q}_{1}+\hat{Q}_{1}\hat{Q}_{1}+\hat{Q}_{2}\hat{Q}+\hat{Q}\hat{Q}_{2}+\cdots)\hat{Q}\hat{Q}^{-1}\\ & \times (\chi_{0}+\chi_{1}+\chi_{2}+\cdots)
    \\  =& H_{\text{eff}}\hat{Q}^{-1}(\chi_{0}+\chi_{1}+\chi_{2}+\cdots),
\end{aligned}
\end{equation}
where $\chi_{n}$ are related to $\hat{\Theta}$ box, $\hat{Q}$ box and their derivatives as
\begin{equation}
\label{chi}
\begin{aligned}
\chi_{0} = &(\hat{\Theta}_{0}+\text{h.c.})+\hat{\Theta}_{00}\\
\chi_{1} = &(\hat{\Theta}_{1}\hat{Q}+\text{h.c.})+(\hat{\Theta}_{01}\hat{Q}+\text{h.c.})\\
\chi_{2} = & (\hat{\Theta}_{1}\hat{Q}_{1}\hat{Q}+\text{h.c.})+(\hat{\Theta}_{2}\hat{Q}\hat{Q}+\text{h.c.}) \\&+ (\hat{\Theta}_{02}\hat{Q}\hat{Q}+\text{h.c.})+\hat{Q}\hat{\Theta}_{11}\hat{Q}.
\end{aligned}
\end{equation}
This perturbation technique of constructing valence-space effective operators of observables has been successful in real-energy SM calculations~\cite{10.3389/fphy.2020.00345, 10.1143/ptp/93.5.905}. In the present work, we apply the MBPT technique of effective operators to the complex GSM in which resonance and continuum states are included. 

In our calculations, the $\chi_{n}$ series is truncated up to the $\chi_{2}$ order, which has been proved to be sufficient to obtain convergences~\cite{10.3389/fphy.2020.00345}. The $\hat{\Theta}$-box diagrams are calculated up to the third order, which is consistent with the expansions used in the $\hat{S}$-box and $\hat{Q}$-box calculations. Due to the presence of the nonresonant continuum, the matrix dimension grows dramatically when adding more valence particles into the continuum. Therefore, we allow at most two valence particles in the continuum, with which converged results can be obtained~\cite{Z.H.Sun-Phys.Lett.B.769.227(2017),HU2020135206,PhysRevC.70.064313}.

In this work, we focus on the GT $\beta$ decay for which the free-space bare transition operator is~\cite{BROWN1985347}
\begin{equation}
    \mathcal{O}(\text{GT}_{\pm})=\sum_{j}\sigma^j\tau^j_\pm,
\end{equation}
which is a one-body operator with $\sigma$ for the Pauli spin operator and $\tau_\pm=(\tau_x\pm i\tau_y)/2$ for isospin operators corresponding to $\beta^\pm$ decays, respectively. The sum is over all interacting nucleons of the nucleus.
Then, the reduced GT transition matrix element~\cite{BROWN1985347} can be calculated by the valence-space wave functions of the initial $|\psi_i\rangle$ and final $|\psi_f\rangle$ states, as
\begin{equation}
    M_\text{GT}=\sum_{{p,q}\in\text{valence space}}M_\text{GT}^{pq}\braket{\psi_{f}\|[\hat{a}_{p}^{\dagger}\hat{a}_{q}]\|\psi_{i}},
\label{M_GT_VS}
\end{equation}
where $M_\text{GT}^{pq}$ stands for valence-space effective GT transition matrix elements obtained with the bare GT operator using the $\Theta$-box perturbation up to the third order. 


In GT transition calculations, usually a quenching factor is needed to better describe data~\cite{BROWN1985347,TOWNER1987263}. 
For the {\it sd} and {\it pf} shells, an average phenomenological quenching factor of $q\approx 0.75$ is usually taken~\cite{BROWN1985347,TOWNER1987263}.
As commented in~\cite{nature.phys.15.428.,PhysRevD.102.074018}, in the chiral EFT framework the effect of the coupling of weak interactions to two nucleons can be calculated via two-body currents, giving a quenching factor as~\cite{PhysRevD.102.074018}
\begin{equation}
\begin{aligned}
    &q = 1 - \frac{\rho}{F_\pi^2}\bigg[\frac{c_4}{3}\left[3 I_2^\sigma(\rho,|\mathbf{q}|)-I_1^\sigma(\rho,|\mathbf{q}|)\right]
    \\&-\frac{1}{3}\left(c_3-\frac{1}{4 m_N}\right) I_1^\sigma(\rho,|\mathbf{q}|)-\frac{c_6}{12} I_{c 6}(\rho,|\mathbf{q}|)-\frac{c_D}{4 g_A \Lambda_\chi}\bigg],
\end{aligned}
\end{equation}
where the momentum transfer $\textbf{q}$ is approximately equal to zero in $\beta$ decays~\cite{PhysRevD.102.074018}. $F_\pi$ and $g_\text{A}$ are the pion-decay and axial-vector coupling constants, respectively. $c_3$, $c_4$, $c_6$ and $c_\text{D}$ belong to the low-energy constants of the EFT with $\Lambda_\chi$ being the chiral scale. These constants have been available in the chiral EM1.8/2.0 interaction~\cite{PhysRevC.83.031301}. Functions $I_1^\sigma(\rho,|\mathbf{q}|)$, $I_2^\sigma(\rho,|\mathbf{q}|)$ and $I_{c 6}(\rho,|\mathbf{q}|)$ have been defined in Refs.~\cite{PhysRevD.88.083516,PhysRevD.89.029901}. The calculated value of the quenching factor is dependent on the nuclear density $\rho$. For finite nuclei, the typical density range is $\rho =0.09-0.11$ $\rm fm^{-3}$~\cite{PhysRevD.102.074018}, and $\rho =0.10$ $\rm fm^{-3}$ is taken usually~\cite{PhysRevD.102.074018}, giving a quenching factor of $q=0.78$ which is used in our calculations. 



The recent experiment~\cite{PhysRevLett.125.192503} observed remarkable isospin asymmetry in mirror nuclei $^{22}\rm Si$-$^{22}\rm O$, and also in their $\beta$-decay mirror daughters $^{22}\rm Al$-$^{22}\rm F$. Two $1^+$ excited states in $^{22}\rm Al$ were observed in 
$\beta$-delayed proton emissions from $^{22}\rm Si$ through $^{22}\rm Al$ to $^{21}\rm Mg$. Isospin asymmetry in mirror nuclei $^{22}\rm Al$-$^{22}\rm F$ was shown by the mirror energy difference (MED)~\cite{PhysRevLett.125.192503}. The MED is a good probe into the mirror isospin asymmetry, which is defined by the difference between the excitation energies of analog states in mirror nuclei. For heavier nuclei in the $pf$ shell, the MED between mirror states is rather small~\cite{BENTLEY2007497,QI200848}. The situation is different for lighter nuclei, such as the $1/2_1^+$ state in $^{13}\rm N$ is 0.72 MeV lower than that in $^{13}\rm C$~\cite{NNDCensdf}, which is related to the loosely bound nature of the proton $1s_{1/2}$ orbit~\cite{PhysRev.88.1109,PhysRev.81.412}. The lack of a centrifugal barrier implies that the radial wave function of the $1s_{1/2}$ orbital has a larger extent than those of other orbitals, which then provides a large MED. The GSM can generate the proper asymptotic behavior of wave functions self-consistently. 

\begin{figure}[h]
\centering
\includegraphics[width=0.45\textwidth]{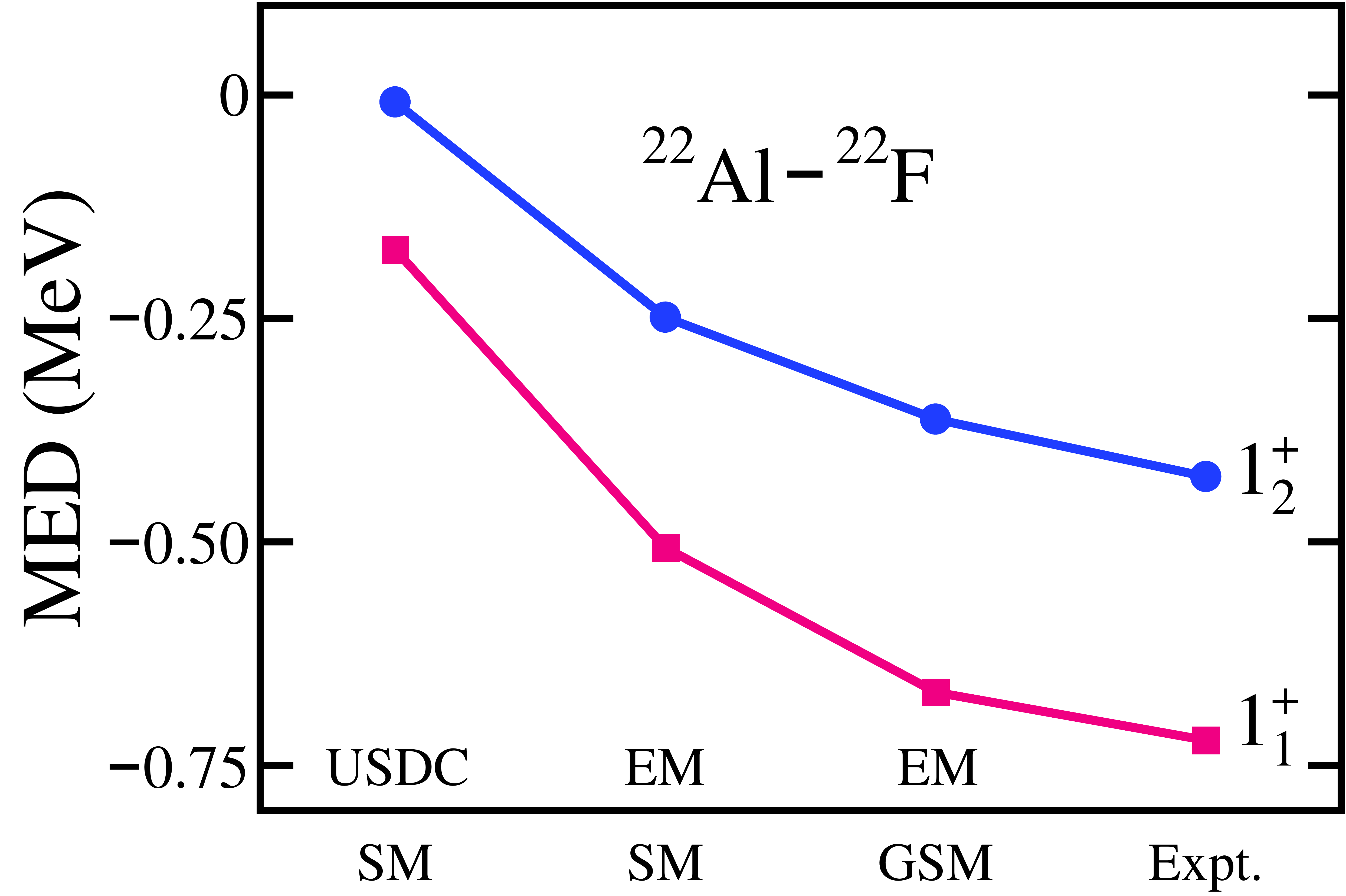}
\caption{MED's for the $1^+_1$ and $1^+_2$ analog states between mirror nuclei $^{22}\rm Al$ and $^{22}\rm F$, calculated by standard SM with USDC and EM1.8/2.0 (abbreviated by EM), and by GSM with EM1.8/2.0, compared with 
data~\cite{PhysRevLett.125.192503}.}
\label{MeD}
\end{figure}

Figure~\ref{MeD} presents calculated and experimental MED's for the first and second $1^+$ analog states between mirror nuclei $^{22}\rm Al$ and $^{22}\rm F$. To see the continuum effect, we have also performed a standard SM calculation using the same EM1.8/2.0 interaction as in the GSM calculation. We see that the inclusion of the continuum coupling improves significantly the MED calculations for both $1^+$ states compared with data. In the experimental paper~\cite{PhysRevLett.125.192503}, the authors explained the observed isospin asymmetry by the standard SM calculation with readjusting isospin-nonconserving (INC) interactions related to the $s_{1/2}$ orbit. The $s_{1/2}$ orbit has no centrifugal barrier, and thus has a strong coupling to the continuum when weakly bound. Both isospin symmetry breaking and coupling to the continuum may be equivalently included by adjusting the relevant interaction matrix elements. As shown in Fig.~\ref{MeD}, we have also performed a standard SM calculation with the new isospin-breaking USD-type interaction, named USDC~\cite{PhysRevC.101.064312} which was obtained by a global fit to {\it sd}-shell data. The result gives smaller MED's than data, indicating the need of the coupling to the continuum. 

\begin{figure}[h]
\centering
\includegraphics[width=0.45\textwidth]
{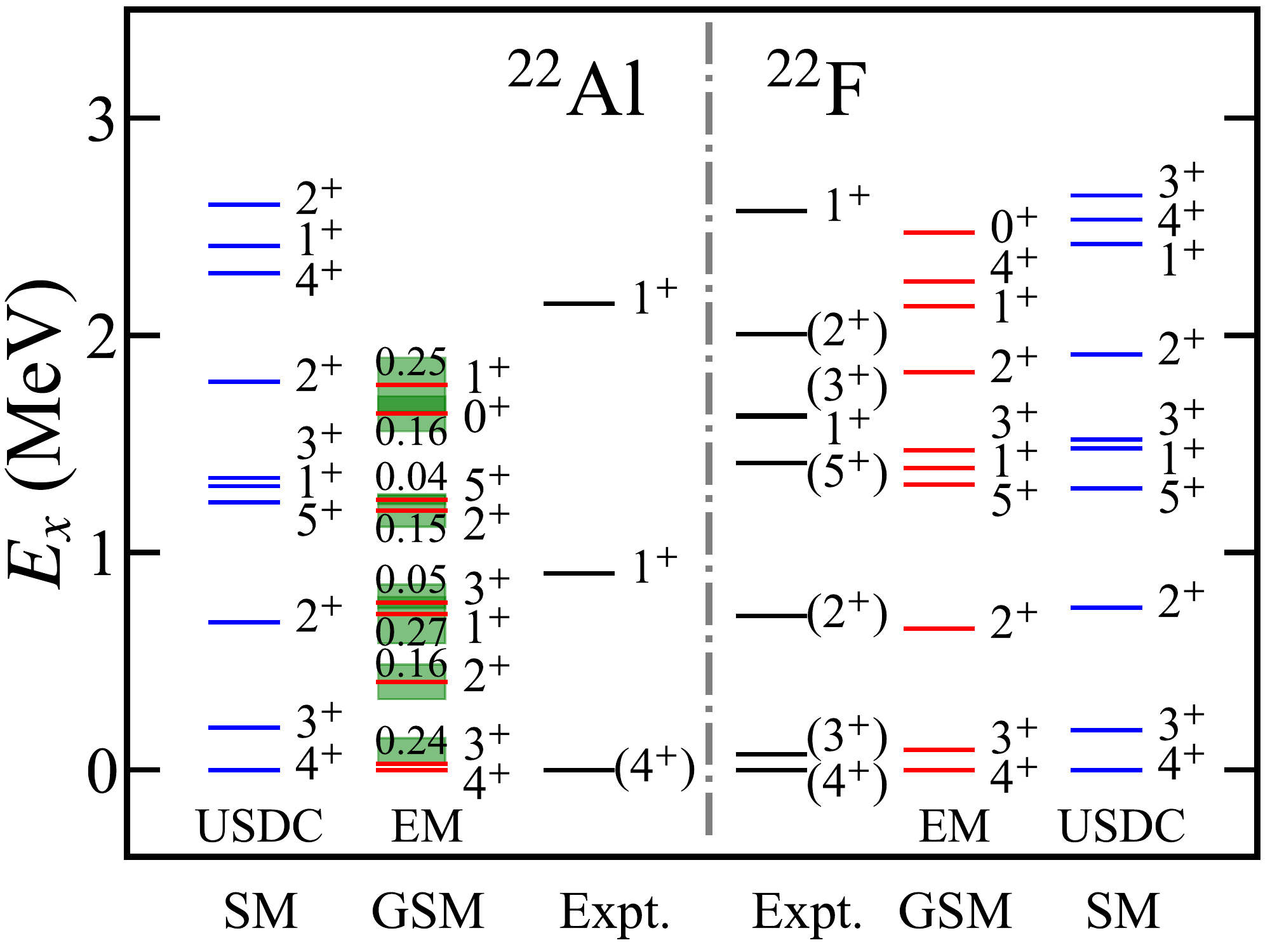} 
\caption{\label{spectra} Spectra of mirror nuclei $^{22}\rm Al$ and $^{22}\rm F$, calculated by GSM with EM1.8/2.0 and by standard SM with USDC. Green shadowing indicates resonance with the calculated width (in MeV) given above (below) the level. The data are taken from~\cite{PhysRevLett.125.192503} for $^{22}\rm Al$ and~\cite{NNDCensdf} for $^{22}\rm F$.}
\end{figure}

Figure~\ref{spectra} shows the spectra of low-lying states for mirror nuclei $^{22}\rm Al$ and $^{22}\rm F$. We see that both GSM with EM1.8/2.0 and standard SM with USDC give reasonable spectra compared with data. The excited states of the proton dripline nucleus $^{22}\rm Al$ should be unbound since the single-proton separation energy is close to zero~\cite{Wang_2021}. In contrast, the single-neutron separation energy in its mirror nucleus $^{22}\rm F$ is 5.230 MeV. Indeed, our GSM calculation predicts that all excited states in $^{22}\rm Al$ are resonances, while $^{22}\rm F$ has bound low-lying states, as shown in Fig.~\ref{spectra}. The experiment~\cite{PhysRevLett.125.192503} detected proton emissions from the two $1^+$ states in $^{22}\rm Al$, indicating the resonances of the $1^+$ states. Both data and calculations show significant TES in the observed $1_1^+$ and $1_2^+$ states between mirror nuclei $^{22}\rm Al$ and $^{22}\rm F$, which has already been shown in Fig.~\ref{MeD} by MED. Besides the $1^+$ levels, the GSM calculations in Fig.~\ref{spectra} also predict the TES phenomenon appearing in other low-spin states in which the $s_{1/2}$ component is significant.

%

The experiment~\cite{PhysRevLett.125.192503} has also observed the mirror asymmetry in the GT $\beta$ decays from $^{22}$Si and $^{22}$O into the $1_1^+$ states of their daughters $^{22}$Al and $^{22}$F. The standard SM calculation can reproduce the $\beta$-decay asymmetry by adjusting the INC interaction matrix elements related to the $s_{1/2}$ orbit~\cite{PhysRevLett.125.192503}. 
The GT transition matrix element is calculated by the wave functions of initial and final states. Therefore, one should properly handle the asymptotic behavior of the weakly bound or quasi-bound states. Indeed, the GSM can well provide valence-space wave functions with resonance and continuum coupling taken into account. 

The calculated magnitude of the GT transition matrix element, $|M_\text{GT}|$, is given in Table~\ref{table1} for the decays of the mirror nuclei $^{22}\rm Si$ and $^{22}\rm O$ into the first and second $1^+$ states of their daughters $^{22}$Al and $^{22}$F. In both nuclei, the calculated $|M_\text{GT}|$ decaying into the $1^+_1$ state is significantly smaller than that into the $1^+_2$ state, which is consistent with data. The isospin asymmetry in the GT transitions into the $1_{1}^{+}$ states is indeed seen in both SM and GSM calculations with the EM1.8/2.0 interaction. Similar to data, 
the $|M_\text{GT}|$ value in $^{22}{\rm O}\rightarrow{^{22}}{\rm F}(1_1^+)$ is almost twice as large as that in $^{22}{\rm Si}\rightarrow{^{22}}{\rm Al}(1_1^+)$ , while this does not happen in the decays into the $1_2^+$ states, as shown Table~\ref{table1}. The GSM calculation with the continuum effect considered improves the agreement with data. The last column of Table~\ref{table1} gives the theoretical $|M_\text{GT}|$ values given in Ref.~\cite{PhysRevLett.125.192503}, obtained by the standard SM with adjusting INC interactions. We have also run a standard SM calculation with USDC, showing excellent agreement with data for the decays into the $1^+_2$ sates, see Table~\ref{table1}. However, the calculation gives too small $|M_\text{GT}|$ value for the $^{22}$O decay into the $^{22}$F $1^+_1$ state to correctly reproduce the asymmetry in the decays to the $1^+_1$ states.
\begin{table}[]
\centering
    \caption{GT transition matrix elements $|M_{\rm GT}|$ calculated by standard SM with USDC and EM1.8/2.0, and by GSM with EM1.8/2.0 for mirror nuclei $^{22}\rm Si$ and $^{22}\rm O$ decaying into the $1_1^+$ and $1_2^+$ states of their mirror daughters $^{22}\text{Al}$ and $^{22}\text{F}$, compared with data and calculations given in Ref.~\cite{PhysRevLett.125.192503}.}
    \label{table1}
    \begin{tabular*}{\hsize}{@{}@{\extracolsep{\fill}}cccccccc@{}}
            \hline\hline 
            & &  \multicolumn{2}{c}{SM} &   \multicolumn{2}{c}{GSM}& \multicolumn{2}{c}{Ref.~\cite{PhysRevLett.125.192503}}\\
            \cline{3-4} \cline{5-6} \cline{7-8}
            & &  USDC  &  EM  &   \multicolumn{2}{c}{EM}& Expt. & Cal. \\
            \hline
         \multirow{2}{*}{$^{22}\text{Si}\to^{22}\text{Al}$}&$1_{1}^{+}$ &  0.236  & 0.343 &  \multicolumn{2}{c}{0.257}& 0.176(16)& 0.242 \\
         &$1_{2}^{+}$ & 0.721 & 1.042 & \multicolumn{2}{c}{1.012}& 0.750(41)& 0.863 \\
         \hline
         \multirow{2}{*}{$^{22}\text{O}\to^{22}\text{F}$}&$1_{1}^{+}$& 0.198 & 0.569 & \multicolumn{2}{c}{0.497}& 0.310(32)& 0.428 \\
         &$1_{2}^{+}$ & 0.719 & 1.092 & \multicolumn{2}{c}{1.068}& 0.775(77)& 0.848\\\hline
         \hline
    \end{tabular*}
\end{table}


To further understand the isospin asymmetry in the GT transitions, we compare the configurations of the states. The SM calculation with EM1.8/2.0 shows that the initial state in the mother nucleus $^{22}\rm Si$ has a dominant configuration with the proton occupation number 5.40 (totaling six valence protons) in the $\pi 0d_{5/2}$ orbital. According to the selection rule of the GT transition, a large occupation in $d$ orbitals in the daughter $^{22}\rm Al$ leads to a large transition matrix element. The SM  calculation shows that the $1^{+}_{1}$ state in $^{22}\rm Al$ has the proton occupation (3.66, 1.18, 0.16) and neutron occupation (0.85, 0.13, 0.02) in ($0d_{5/2}$, $1s_{1/2}$, $0d_{3/2}$) respectively. Conversely, the $1^{+}_{2}$ state in $^{22}\rm Al$ has the proton occupation (4.26, 0.35, 0.40) and neutron occupation (0.77, 0.16, 0.07) in  ($0d_{5/2}$, $1s_{1/2}$, $0d_{3/2}$) respectively. The main difference between the two $1^{+}$ states then clearly appears: the $1^{+}_{1}$ state has a large proton occupation in $1s_{1/2}$ but a small proton occupation in $0d_{3/2}$. In contrast, the $1^{+}_{2}$ state has a reduced proton $1s_{1/2}$ occupation but an enhanced proton $0d_{3/2}$ occupation. Therefore, the transition density and the matrix element of the GT transition into the  $1^{+}_{2}$ state are larger than those into the $1^+_1$ state because of a larger occupation in $d$ orbitals for the $1^{+}_{2}$ state. This result is consistent with the data, as shown in Table~\ref{table1}. The situation is similar in the SM calculation with USDC. The GSM calculation with EM1.8/2.0 also gives similar results but a further enhanced proton $1s_{1/2}$ occupation of 1.39 in the $1_1^+$ state of $^{22}\rm Al$, which is from the continuum coupling. However, the enhancement is not seen in the mirror $\beta^-$ transition in which the $\nu 1s_{1/2}$ occupation in the $^{22}\rm F$ $1_1^+$ state is 1.11. As the continuum coupling improves the asymptotic behavior of the resonant $\pi 1s_{1/2}$ orbital, a better description of the large difference in $M_\text{GT}$
 between the decays into those two $1_1^+$ states of the mirror nuclei $^{22}\rm Al$-$^{22}\rm F$ is obtained in the GSM calculation. 

\begin{figure}[h]
\centering
\includegraphics[width=0.5
\textwidth]
{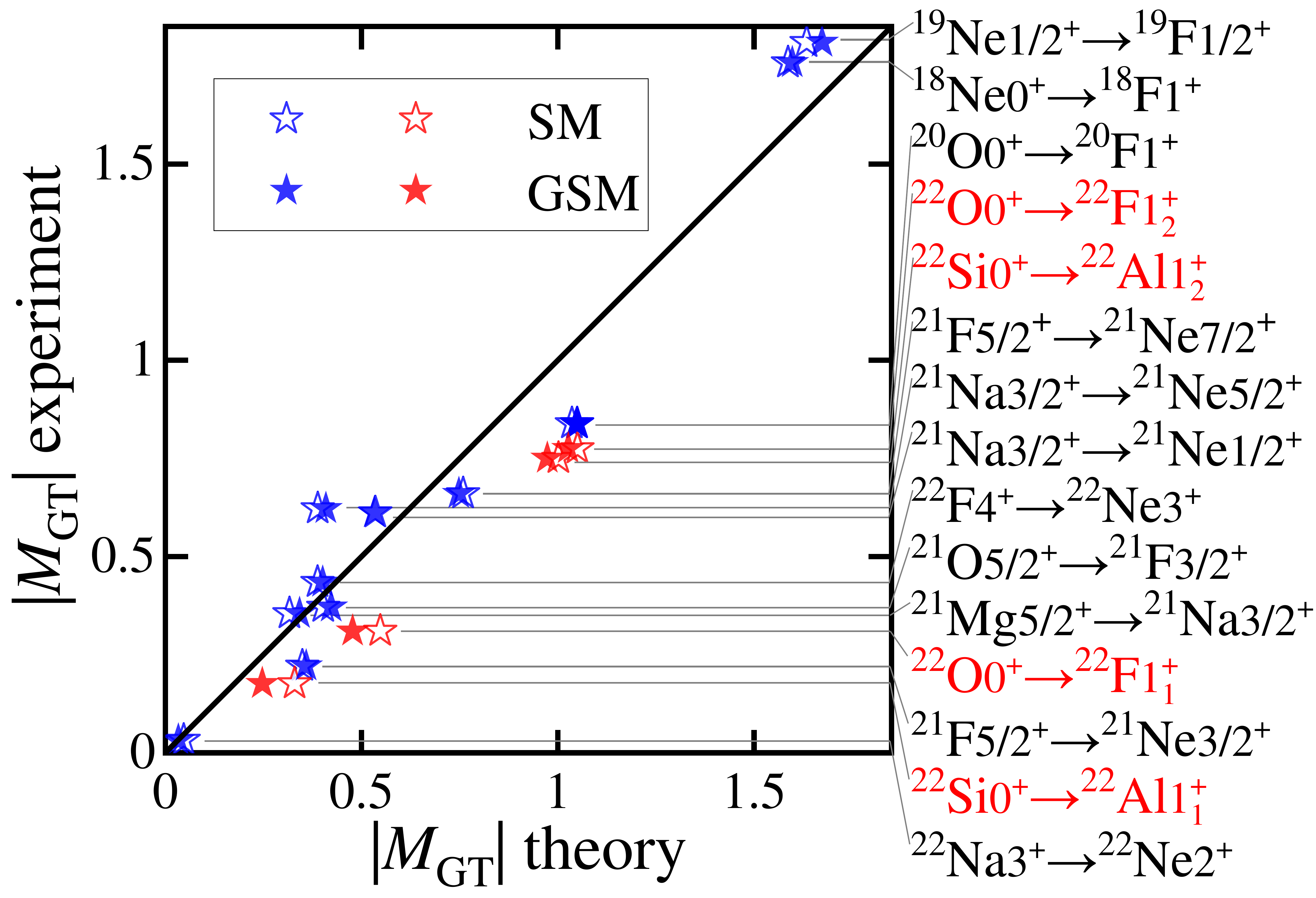}
\caption{\label{M_GT_all}$A\approx 20$ GT transition matrix elements calculated by GSM and SM with EM1.8/2.0, compared with data~\cite{PhysRevLett.125.192503,NNDCensdf}.}
\end{figure}

We have systematically investigated the GT transitions for $A \approx 20$ nuclei, as shown in Fig.~\ref{M_GT_all}. Calculated $|M_{\rm GT}|$ matrix elements are in reasonable agreement with data. For nuclei near the $\beta$ stability, the $M_{\rm GT}$ values by GSM are almost the same as those by SM. This should be expected because the continuum effect in stable mass regions can be neglected. However, for exotic nuclei far from the $\beta$ stability, one can observe that the continuum effect can be visible in the calculations of $M_{\rm GT}$ and isospin asymmetry, e.g., for the mirror nuclei $^{22}\rm Si$-$^{22}\rm O$ and their mirror daughters $^{22}\rm Al$-$^{22}\rm F$. As seen from Fig.~\ref{M_GT_all}, the experimental $|M_{\rm GT}|$ values are almost the same for the $^{22}\rm Si$ and $^{22}\rm O$ $\beta$ decays whose final state is $1_{2}^{+}$. However, in the case of the $1_1^+$ final state, the $|M_{\rm GT}|$ values between the mirror nuclei differ, thus revealing the isospin asymmetry. Also, 
one can notice that the continuum coupling evidently improves the calculations of the $|M_{\rm GT}|$ decaying into the $1_{1}^{+}$ states in the mirror transitions $^{22}{\rm Si}\rightarrow{^{22}}{\rm Al}$ and $^{22}{\rm O}\rightarrow{^{22}}{\rm F}$. 

In summary, we have extended the $\it ab\ initio$ many-body perturbation theory to the complex Berggren basis to devise valence-space effective operators for the Gamow shell model which takes the continuum coupling into account. The Berggren basis is generated by the Gamow Hartree-Fock method with the same realistic interaction used as in the Gamow shell-model calculation, which leads to the self-consistent treatments of spectra and other observables within the MBPT framework. In the presence or absence of the continuum coupling, we have calculated spectra and Gamow-Teller transition matrix elements for $sd$-shell nuclei. Mirror nuclei $^{22}\rm Al$ and $^{22}\rm F$ exhibit significant isospin asymmetry in the associated GT transitions and spectra, where the $s_{1/2}$ partial wave plays a crucial role. 

As expected, the continuum coupling barely affects the GT transitions of nuclei close to the $\beta$ stability. However, the continuum coupling is necessary to properly describe the spectra and GT transitions of nuclei in the vicinity of driplines. A suitable mirror energy difference between $^{22} \text{Al}$ and $^{22} \text{F}$ can be provided by the calculation with the continuum coupling present. The Thomas-Ehrman shift can be explained by the continuum effect which increases the occupation of the $s_{1/2}$ orbital. 

The decay into the $1^+_2$ state has a larger GT matrix element than that into the $1^+_1$ state in the mirror transitions $^{22}{\rm Si}\rightarrow{^{22}}{\rm Al}$ and $^{22}{\rm O}\rightarrow{^{22}}{\rm F}$. This results from the larger $d$-orbital occupation in the $1^+_2$ state, which increases the $|M_{\rm GT}|$ value. The resonant proton $1s_{1/2}$ orbital is crucial for a proper description of observables related to the $1_1^+$ excited state in $^{22}\text{Al}$. The absence of a centrifugal barrier for the $1s_{1/2}$ orbital results in an extended wave function in the coordinate space, 
and hence a lower $|M_{\rm GT}|$ value. The proper asymptotic behavior of the quasi-bound $1^+_1$ state in  $^{22}\rm Al$ can then provide a better description of the GT transition.

This work has been supported by the National Natural Science Foundation of China under Grants No. 12335007, 11835001, 11921006, 12035001, 12105106, 12205340, and 12175281; 
Gansu Natural Science Foundation under Grant No. 22JR5RA123.
\normalem
\bibliography{article}

\end{document}